\PassOptionsToPackage{table}{xcolor}

\documentclass[a4paper,fleqn,usenatbib]{mnras_tex_edited}

\usepackage{amssymb,amsmath,verbatim,mathtools,needspace,enumitem,etoolbox,graphicx,physics,microtype,natbib,url,hyperref,mathrsfs,bm,ulem}

\newcommand{\ssim}{\mathchar"5218\relax\,}

\newcommand{\bham}{School of Physics and Astronomy and Institute for Gravitational Wave Astronomy, University of Birmingham,\vspace{-0.05cm}\\$\;$Birmingham, B15 2TT, UK
}

\renewcommand{\emph}[1]{\textit{#1}}

\def\apj{\rm{ApJ}}
\def\apjl{\rm{ApJ}}
\def\apjs{\rm{ApJS}}
\def\aap{\rm{A\&A}}

\def\mnras{\rm{MNRAS}}
\def\prd{\rm{PRD}}
\def\prl{\rm{PRL}}
\def\prx{\rm{PRX}}

\def\cqg{\rm{CQG}}
\def\lrr{\rm{LRR}}

\title[Are stellar-mass BH binaries too quiet for LISA?]{Are stellar-mass black-hole binaries too quiet for LISA?
}

\author[Moore et al.]{
Christopher J. Moore\thanks{\href{mailto:cmoore@star.sr.bham.ac.uk}{cmoore@star.sr.bham.ac.uk}},
Davide Gerosa,
Antoine Klein
\bigskip \\
\bham
\vspace{-0.6cm}
}
\pubyear{2019}

\begin{document}

\label{firstpage}
\pagerange{\pageref{firstpage}--\pageref{lastpage}}
\maketitle

\begin{abstract}
The progenitors of the high-mass black-hole mergers observed by LIGO and Virgo are potential LISA sources and promising candidates for multiband GW observations.
In this letter, we consider the minimum signal-to-noise ratio these sources must have to be detected by LISA bearing in mind the long duration and complexity of the signals. Our revised threshold of $\rho_{\rm thr}\ssim 15$ is higher than previous estimates, which significantly reduces the expected number of events. 
We also point out the importance of the detector performance at high-frequencies and the duration of the LISA mission, which both influence the event rate substantially.
\end{abstract}


\section{Introduction}

Until recently, stellar-mass black holes (BHs) were only thought to be as heavy as $\ssim 10 M_\odot$, as inferred from X-ray binary measurements (e.g.\ \citealt{2014bsee.confE..37W}).
The first detections of gravitational-waves (GWs) by the LIGO/Virgo detectors point to a population of BHs with masses up to $\ssim 30 M_\odot$ \citep{2018arXiv181112907T}.
This difference has important consequences for the  formation and evolution of BH binaries-- for instance, proving that low-metallicity environments play a vital role \citep{2010ApJ...714.1217B,2016ApJ...818L..22A}.
It is also crucial for the future of GW astronomy.

BH binaries with components of $\ssim30 M_{\odot}$ might emit GWs strongly enough at mHz frequencies to be within reach of LISA~\citep{2016PhRvL.116w1102S}.
This opens up the exciting possibility of performing \emph{multiband} GW astronomy: a single source being observed by both LISA and  LIGO.
Following this realisation, stellar-mass BH binaries started to be explored as an important part of the LISA science case, in terms of astrophysics~\citep{2010ApJ...725..816B,
2016ApJ...830L..18B,
2016PhRvD..94f4020N,2017MNRAS.465.4375N,
2016MNRAS.462.2177K,
2016PhRvL.116w1102S,
2017JPhCS.840a2018S,
2018arXiv180208654S,
2018MNRAS.481.4775D,
2018MNRAS.481.5445S,2019PhRvD..99f3006S,
2018PhRvL.120s1103K,2019PhRvD..99f3003K,
2019ApJ...878...75R,
2019ApJ...875...75F,
2019PhRvD..99j3004G}, fundamental physics \citep{
2016PhRvL.116x1104B,2017PhRvD..96h4039C,2018arXiv180700075T}, cosmology~\citep{2017PhRvD..95h3525K,2018MNRAS.475.3485D}, and data analysis~\citep{2016PhRvL.117e1102V,2018PhRvL.121y1102W,2019PhRvD..99f4056M,2019BAAS...51c.109C,2019arXiv190508811T} 

In this letter, we add a point of caution.
Stellar-mass BH binaries can emit in the LISA band for the entire duration of the mission, generating millions of GW cycles with a complex, chirping signal morphology.
These will need to be extracted from the LISA datastream.
If this were to be done using templates, we estimate the size of the template bank required and, consequently, the threshold signal-to-noise ratio (SNR) where events are loud enough to be detected.
We find 
an expected SNR threshold  $\rho_{\rm thr} \ssim 15$ for systems merging within 10~yr.
Previous work has assumed a threshold $\rho_{\rm thr} \ssim 8$, similar to that for compact-binary mergers in LIGO~\citep{2016ApJ...833L...1A}, 
and predict 
a handful of multiband detections
~\citep{2016MNRAS.462.2177K,2017JPhCS.840a2018S,2019PhRvD..99j3004G}.
The expected number of events 
scales as  $\rho_{\rm thr}^{-3}$. Therefore, increasing the SNR threshold by nearly a factor of $\ssim 2$ 
has important consequences.
Stellar-mass BH binaries might just
 be too quiet for LISA.

Even if searching for stellar-mass BHs directly in LISA data turns out to be difficult, some of these signals could be extracted from the noise \emph{a posteriori} \citep{2018PhRvL.121y1102W}. Detections from ground-based interferometers will allow us to revisit past LISA data hunting for signals with known parameters. In this case, we find $\rho_{\rm thr}\ssim 9$.

The rest of this letter elaborates on these findings. In Sec.~\ref{sec:rho_threshold}, we relate the threshold SNR to the size of the template bank. The required density of templates is estimated in Sec.~\ref{sec:Nbank} with a Fisher-matrix calculation. In Sec.~\ref{sec:implications} we estimate LISA detection rates. In Sec.~\ref{sec:conclusions} we draw our conclusions . 
We assume cosmological parameters from \cite{2016A&A...594A..13P}.

\section{The Threshold SNR} \label{sec:rho_threshold}

GW detection is routinely performed using template banks.
These searches involve matching sets of precomputed waveform templates against the observed data.
The threshold SNR above which a signal can be confidently detected depends on the number of templates in the bank.
This relationship is derived for an idealised search in this section using methods similar to those in \citet{2003PhRvD..67b4016B} and \cite{2017PhRvD..96d4005C}.

Let us assume that the GW signals may be written as
\begin{equation}
h_\alpha(t)=\rho\, \hat{h}_\alpha(t)\exp(i\phi_s),
\end{equation}
where the $\hat{h}_\alpha$ are the normalised template waveforms with $| \hat{h}_\alpha |=1$.
A hypothetical template bank $\{\hat{h}_\alpha\,|\,\alpha=1,2,\ldots,N_{\rm bank}\}$ may be constructed spanning all the source parameters except for the SNR, $\rho$, and the phase shift, $\phi_s$, which may be searched over for each $\hat{h}_\alpha$ at negligible additional cost.
In practice one would not usually use a template bank to search over the time-of-arrival parameter, as this can be handled more efficiently using fast Fourier transform techniques \citep{1998PhRvD..57.2101B}; however, for our hypothetical search it is convenient to imagine treating this the same as the other parameters.

The detection statistics are the phase-maximised inner product between the data, $s$, and the templates,
\begin{equation}
\sigma_\alpha=\max_{\phi_s}\,\langle s|\hat{h}_\alpha\rangle.
\end{equation}
    When the data contains only stationary, Gaussian noise ($s=n$) the statistics $\sigma_\alpha$ follow a Rayleigh distribution with probability density
\begin{equation}
f_0(\sigma_\alpha)=\sigma_\alpha\exp\left(-\frac{\sigma_\alpha^2}{2}\right),
\end{equation}
with $\sigma_\alpha\geq0$. If a signal is present ($s=h_\alpha+n$), the statistic for the corresponding template follows a Rice distribution with offset $\rho$. This has probability density
\begin{equation}
f_1(\sigma_\alpha,\rho)=\sigma_\alpha\exp\left(-\frac{\sigma_\alpha^2+\rho^2}{2}\right) I_0(\rho\sigma_\alpha),
\end{equation}
where $I_0$ denotes the zeroth-order modified Bessel function of the first kind. 

A detection is claimed if at least one of the $\sigma_\alpha$ exceed a predetermined threshold, $\sigma_\mathrm{thr}$.
This threshold is set by requiring a certain (small) false-alarm probability:
\begin{align}
&P_F(\sigma_\mathrm{thr})=\int_{\sigma_\mathrm{thr}}^\infty \!\!\mathrm{d}\sigma_\alpha\;f_0(\sigma_\alpha)
\;\Rightarrow\;\sigma_\mathrm{thr}(P_F)\!=\!\sqrt{-2\ln{P_F}}.
\end{align}
A typical\footnote{
For example, \cite{2016PhRvX...6d1015A} use a false-alarm rate threshold of $\mathrm{FAR}=0.01\,\mathrm{yr}^{-1}$ for an observation period of $T=51.5\,\mathrm{days}$. This corresponds to $P_F=1-\mathrm{e}^{-T\,\mathrm{FAR}}\approx1.4\times 10^{-3}$.}
choice for $P_F$ across the bank might be $10^{-3}$.
Approximating the statistics $\{\sigma_\alpha\,|\,\alpha=1,2,\ldots,N_{\rm bank}\}$ as independent random variables, the false-alarm probability in a single template is approximately reduced by a factor $N_{\rm bank}$. Hence we set
\begin{equation}\label{eq:fixed_FAR}
P_F=\frac{10^{-3}}{N_{\rm bank}}.
\end{equation}

The detection probability (i.e.\ the probability that, in the presence of a signal, the statistic for the corresponding template exceeds the threshold; $\sigma_\alpha>\sigma_\mathrm{thr}$) is given by
\begin{equation}\label{eq:P_D}
P_D(\rho)=\int_{\sigma_\mathrm{thr}}^\infty \mathrm{d}\sigma_\alpha\,f_1(\sigma_\alpha,\rho) \approx\Theta(\rho-\rho_{\rm thr}) .
\end{equation}
This detection probability rises from zero to unity across a narrow range $\Delta\rho\approx1$ and can be modelled as a Heaviside step function, $\Theta$. 
Here we are assuming that all sources with $\rho>\rho_{\rm thr}$ are recovered whilst all other sources are missed.

\begin{figure}
\centering
\includegraphics[width=0.45\textwidth]{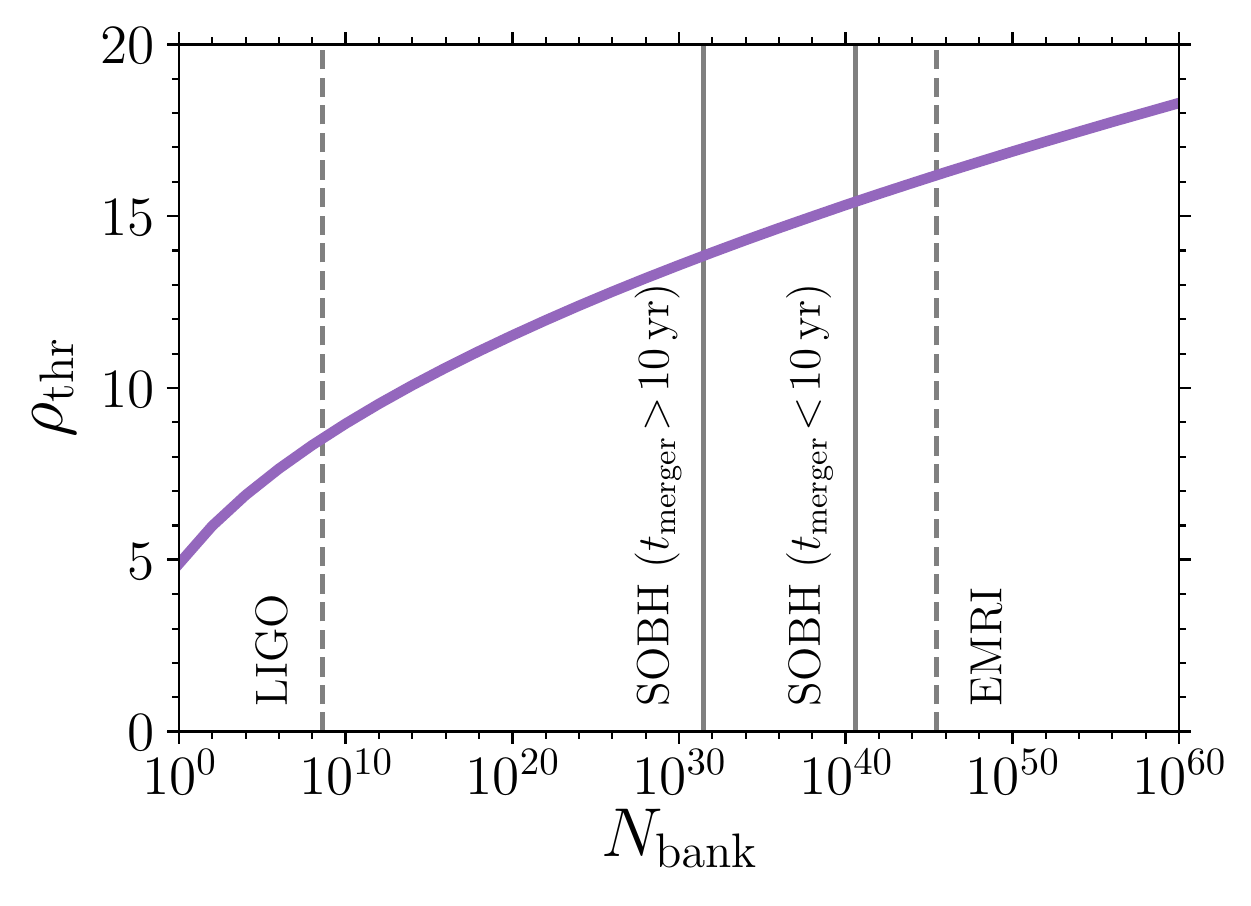}
\caption{\label{fig:thresh}
The threshold SNR as a function of template bank size.
Solid vertical lines indicate the bank sizes for stellar origin black hole (SOBH) binaries with component masses in the range $(5$ - $50)\,M_{\odot}$; these are split into those which merge in $<10\,\mathrm{yrs}$ (fast chirping) and $>10\,\mathrm{yrs}$ (slow chirping).
The thresholds are $\rho_{\rm thr}\ssim 15$ and $\ssim 14$ respectively (see Table.~\ref{tab:rhothresh-vs-mass}).
For comparison, two other classes of GW are also indicated.
Binary BHs in LIGO/Virgo can be detected with single-detector SNRs as low as $\ssim8$ using banks containing $\ssim4\times10^{5}$ templates  (\citealt{2017arXiv170501845D}; this number does not include the time-of-arrival parameter; including this enlarges the effective size of the bank by a factor of $\ssim10^{3}$, the number of cycles in a typical template).
At the other extreme, EMRIs in LISA have $\rho_{\rm thr}\gtrsim16$ and would require 
very large  template banks 
\citep{2017PhRvD..96d4005C}.
As discussed in the text, we do \emph{not} propose to actually use such huge template banks in practical searches; these are estimates of the numbers required by a hypothetical, optimal search and provide lower bounds on the threshold for a practical, possibly suboptimal search.
\vspace{-0.1cm}
}
\end{figure}

The threshold SNR depends on the size of the template bank through the trials factor $N_{\rm bank}$ in Eq.~\eqref{eq:fixed_FAR}; this dependence is plotted in Fig.~\ref{fig:thresh}.

The above discussion considered an idealised template bank search and gave no consideration to computational costs.
Some of the template banks indicated in Fig.~\ref{fig:thresh} are far too large to be practically implemented.
In those cases it is necessary to use an alternative procedure.
For example, when searching for extreme mass-ratio inspirals (EMRIs) in LISA data, a semi-coherent search strategy has been proposed \citep{2004CQGra..21S1595G} that involves splitting the data into segments, searching each segment separately with small template banks, and combining the results into an overall detection statistic.
Multiband binaries might require a similar approach. 
This computationally viable alternative is suboptimal compared to a full template bank search and this can further raise the detection threshold (in the EMRI case, from $\rho_{\rm thr}\ssim 16$ to $\gtrsim 20$; \citealt{2017PhRvD..96d4005C}).
On the other hand, it might be possible to compress the template bank; i.e.\ to accurately cover the signal space of interest with a reduced basis \citep{2010PhRvD..82d4025C, 2011PhRvL.106v1102F}. Such a compression scheme would reduce the effective size of the template bank potentially lowering the detection threshold from that estimated here.
Future work should address the practical implementation of a search and assess the sensitivity via the blind injection and recovery of signals into mock LISA data.

\section{The size of the template bank} \label{sec:Nbank}

Let us now estimate the size of the template bank $N_\mathrm{bank}$ required to detect stellar-mass BH binaries with LISA.
We consider the following parameters
\begin{align}
\lambda^{\mu}
&\! \in \!\{ \ln\!m_1, \ln\!m_2, \cos \theta_N, \phi_N, \cos \theta_L, \phi_L, e_0^2, \phi_e, \chi_\mathrm{eff}, t_{\rm merger} \} \,, \nonumber
\end{align}
where $m_i$ is the mass of object $i$, $\theta_N$ and $\phi_N$ are angles describing the source's sky location, $\theta_L$ and $\phi_L$ are angles describing the direction of the sources orbital angular momentum, $e_0$ is the eccentricity at $t=0$, $\phi_e$ is the argument of periastron at $t=0$, $\chi_\mathrm{eff}$ is the effective spin parameter, and $t_{\rm merger}$ is the time to merger from $t=0$. The LISA mission starts collecting data at $t=0$.

We adopt a conservative approach and do not include spin components other than $\chi_{\rm eff}$ when estimating the size of the bank. If these parameters are significant for a fraction of the source population-- in particular systems with small $t_{\rm merger}$ \citep{2019PhRvD..99f4056M}-- they will provide an additional contribution to the overall size of the bank.

The Fisher matrix
\begin{align}
\Gamma_{\mu\nu} = \left<\frac{\partial\hat{h}}{\partial\lambda^{\mu}}\Big|\frac{\partial\hat{h}}{\partial\lambda^{\nu}}\right> 
\end{align}
provides a natural metric on parameter space to guide where templates should be placed \citep{1996PhRvD..53.6749O,1999PhRvD..60b2002O, 1991PhRvD..44.3819S,1994PhRvD..49.1707D}.
The diagonal components of $\Gamma$ are the squared reciprocals of the natural length scale for the template separation along each parameter direction.
In order to ensure that there is at least one template along each dimension; we employ a modified Fisher matrix
\begin{align}
\tilde{\Gamma}_{\mu\nu} = \mathrm{max}\left( \Gamma_{\mu\nu} , \frac{\delta_{\mu\nu}}{\left[\Delta\lambda^{\mu}\right]^2} \right),
\end{align}
where $\Delta\lambda^{\mu}$ is the prior range on the parameter $\lambda^{\mu}$.
This modification is only important for parameters which have very little effect on the waveform (e.g.\ $\chi_\mathrm{eff}$ for systems far from merger).

The total number of templates in the bank is found by integrating over the parameter space~\citep{2004CQGra..21S1595G,2005CQGra..22S.927C}
\begin{align}
N_{\rm bank} \approx \int\mathrm{d}\lambda\;\sqrt{\mathrm{det}\,\tilde{\Gamma}} \,.
\label{nbanksqrt}
\end{align}
The square root of $\mathrm{det}\,\tilde{\Gamma}$ gives the template number density required such that the mismatch between adjacent templates is $\ssim50\%$.
This mismatch is larger than that used in practical searches \citep{2016CQGra..33u5004U,2017PhRvD..95d2001M}, but here serves to estimate the number of \emph{independent} templates in the bank, as required by Eq.~\eqref{eq:fixed_FAR}.

We evaluate this integral using Monte Carlo integration. We use templates described by~\cite{2018PhRvD..98j4043K}, setting the spins to $\bm{S}_i = m_i^2 \chi_{\mathrm{eff}} \hat{\bm{L}}$. We compute the determinant of the Fisher matrices using the noise curve given by~\cite{2019CQGra..36j5011R}, being careful to remove near-singular matrices. We focus on binaries observed by LIGO/Virgo and set $m_1,m_2 \in [5 M_\odot, 50 M_\odot]$~\citep{2018arXiv181112907T}. We consider both fast-  ($0< t_{\rm merger} < 10$ yr) and slow-chirping ($10$~yr$< t_{\rm merger} < 100$~yr) sources and set a range of eccentricities $0 < e_0 < 0.4$.

Our results are presented in Table~\ref{tab:rhothresh-vs-mass}. 
We find SNR thresholds for fast chirping binaries between $14.9\lesssim \rho_{\rm thr}\lesssim 15.4$. 
Slow chirping sources, on the other hand, are easier to detect; we find $13.5\lesssim \rho_{\rm thr}\lesssim 13.9$. 
The lower (upper) edge of these ranges correspond to heavier (lighter) systems, with fewer (more) cycles in band.
These estimates are significantly higher than the threshold $\rho_{\rm thr}\ssim8$ typically used in the literature.

We stress that the dependency of $\rho_{\rm thr}$ on $N_{\rm bank}$ in Fig.~\ref{fig:thresh} is rather flat. Although tweaking the parameter ranges to be covered by a search changes the number of templates required, it has only a modest impact on the threshold SNR.
    
  \subsection{Archival searches}
     \label{subsec:archival}

Revisiting past LISA data in light of ground-based observations is a promising avenue to detected more events \citep{2018PhRvL.121y1102W}. In such a scenario, the targeted template bank can be restricted given prior knowledge on the source. 
    For concreteness, we consider an archival search corresponding to a GW150914-like event detected by a third generation ground-based detector 4 years after the start of the LISA mission. The integral in Eq.~\eqref{nbanksqrt} is computed restricting its parameter range to the measurements errors of GW150914 \citep{2016PhRvL.116x1102A} reduced by a factor of 10. We also assume a perfect measurement of $t_{\rm merger}$ and do not integrate over  it.
    Prior information from the ground allows to decrease the size of the template bank by a factor of $\ssim 10^{29}$, reducing the threshold to $\rho_{\rm thr}\simeq9.4$ (Table~\ref{tab:rhothresh-vs-mass}).
    
    \cite{2018PhRvL.121y1102W} considered simulated LISA triggers and also found an improvement of a factor of $\ssim 2$ in $\rho_{\rm thr}$ for archival searches. Our results are largely consistent with this improvement factor.
    
    \begin{table}
        \begin{center}
            \begin{tabular}{c|c|c|c|c|c|}
                $m_1,m_2\;[M_\odot]$ & $N_{\mathrm{bank}}^{\rm (fast)}$ & $N_{\mathrm{bank}}^{\rm(slow)}$ & $\rho_{\rm thr}^{\rm(fast)}$ & $\rho_{\rm thr}^{\rm(slow)}$  \\
                \hline
                \rowcolor{lightgray} $5\!-\!50$ & $10^{40.6}$ & $10^{31.5}$  &$15.4$ & $13.9$\\
                $10\!-\!50$ & $10^{38.4}$ & $10^{30.5}$ & $15.1$ & $13.7$ \\
                $20\!-\!50$ & $10^{37.5}$ & $10^{29.8}$ & $14.9$ & $13.5$ \\
                archival & $10^{11.7}$ & -- & $9.4$ & -- \\
            \end{tabular}
        \end{center}
        \caption{\label{tab:rhothresh-vs-mass}Total effective number of templates in the bank and corresponding threshold SNR. We consider  different lower-mass limits, as well as a representative archival search for a GW150914-like event. Superscripts $^{\rm(fast)}$ and $^{\rm(slow)}$ correspond to fast- ($0 < t_{\rm merger} < 10$~yr) and slow-chirping  ($10$~yr~$< t_{\rm merger} < 100$~yr) binaries, respectively. The results for the row highlighted in gray are shown in Fig.~\ref{fig:thresh}.
}
    \end{table}

    \section{Number of Multiband Events} \label{sec:implications}

    We now assess the impact of our revised SNR threshold on a simple, but realistic astrophysical population of stellar-mass BH binaries. Our procedure closely mirrors that of \cite{2019PhRvD..99j3004G}, to which we refer for further details.
    
    The number of multiband detections is estimated by
    \begin{align}
    &N_{\rm multib} \!= \int {\rm d}z \, {\rm d}\zeta\, {\rm d}\theta\, {\rm d}t_{\rm merger}\, \mathcal{R}(z)\;p(\zeta) p(\theta) \frac{{\rm d} V_c(z)}{{\rm d}z} \frac{1}{1+z}\times
    \notag \\
    &\!\Theta[\rho(\zeta,\theta,t_{\rm merger})- \rho_{\rm thr}]
    \; \mathcal{F} \,  p_{\rm det}(\zeta, z)
    \Theta(T_{\rm wait}-t_{\rm merger})\,.
    \label{Nmultibib}
    \end{align}
    Here $\zeta$ collectively denotes BH masses, spins, and binary eccentricity, $p(\zeta)$ is their probability distribution function, $z$ is the redshift, $V_c$ is the  comoving volume,  $\mathcal{R}(z)$ is the intrinsic merger rate density, $\theta$ collectively denotes the angles $\theta_N$, $\phi_N$, $\theta_L$ and $\phi_L$, and $p(\theta)$ is the corresponding probability distribution function.
        
    For simplicity, we consider non-spinning BHs  on quasi-circular orbits, i.e. $\zeta=\{m_1,m_2\}$.
We assume  $m_1,m_2 \in [5 M_\odot, 50 M_\odot]$ distributed according to $p(m_1)\propto m_1^{-2.3}$ and $p(m_2|m_1) =\,$const. For this mass spectrum, \cite{2018arXiv181112907T} measured $\mathcal{R}= 57^{+40}_{-25}\,{\rm Gpc}^{-3} {\rm yr}^{-1}$. We stress that uncertainties in both $\mathcal{R}$ and $p(\zeta)$ affect our predictions.
     
    \cite{2019PhRvD..99j3004G} used a sky-averaged LISA noise curve to compute SNRs. Here we perform a more generic calculation where we compute $\rho$ as a function of $\theta$. This allows us to capture individual events that are expected to be above threshold only for favorable orientations or positions in the sky.  We use the low-frequency LISA response by \cite{1998PhRvD..57.7089C} and  waveforms by \cite{2010PhRvD..82f4016S}. Binaries are distributed uniformly in $\cos\theta_N$, $\phi_N$, $\cos\theta_L$ and $\phi_L$.  The initial frequency is set by $t_{\rm merger}$. LISA SNRs are computed using the mission specification of \cite{2019CQGra..36j5011R} and a mission duration $T_{\rm obs}$ of 4 or 10 yrs. Events are then selected using a cut in SNR at $\rho_{\rm thr}$.
    
    The term
    $p_{\rm det}(\zeta,z)$ in Eq.~\eqref{Nmultibib} encodes selection effects of the ground based detector, and is estimated using the single-detector approximation by \cite{1993PhRvD..47.2198F}. As stressed by \cite{2019PhRvD..99j3004G}, the multiband detection rate is largely independent of the specifications of the ground-based network. For concreteness we assume a LIGO instrument at design sensitivity~\citep{2018LRR....21....3A}, but we have also verified that identical results are obtained if a third generation detector is used instead.
    For multiband scenarios, one might be interested only in binaries merging within a given timeframe $T_{\rm wait}$, and thus limit the detection rate to sources with $t_{\rm merger}<T_{\rm wait}$. For simplicity, we assume a ground-based network with a duty cycle $\mathcal{F}=1$.

\begin{figure}
    \centering
        \includegraphics[width=0.45\textwidth]{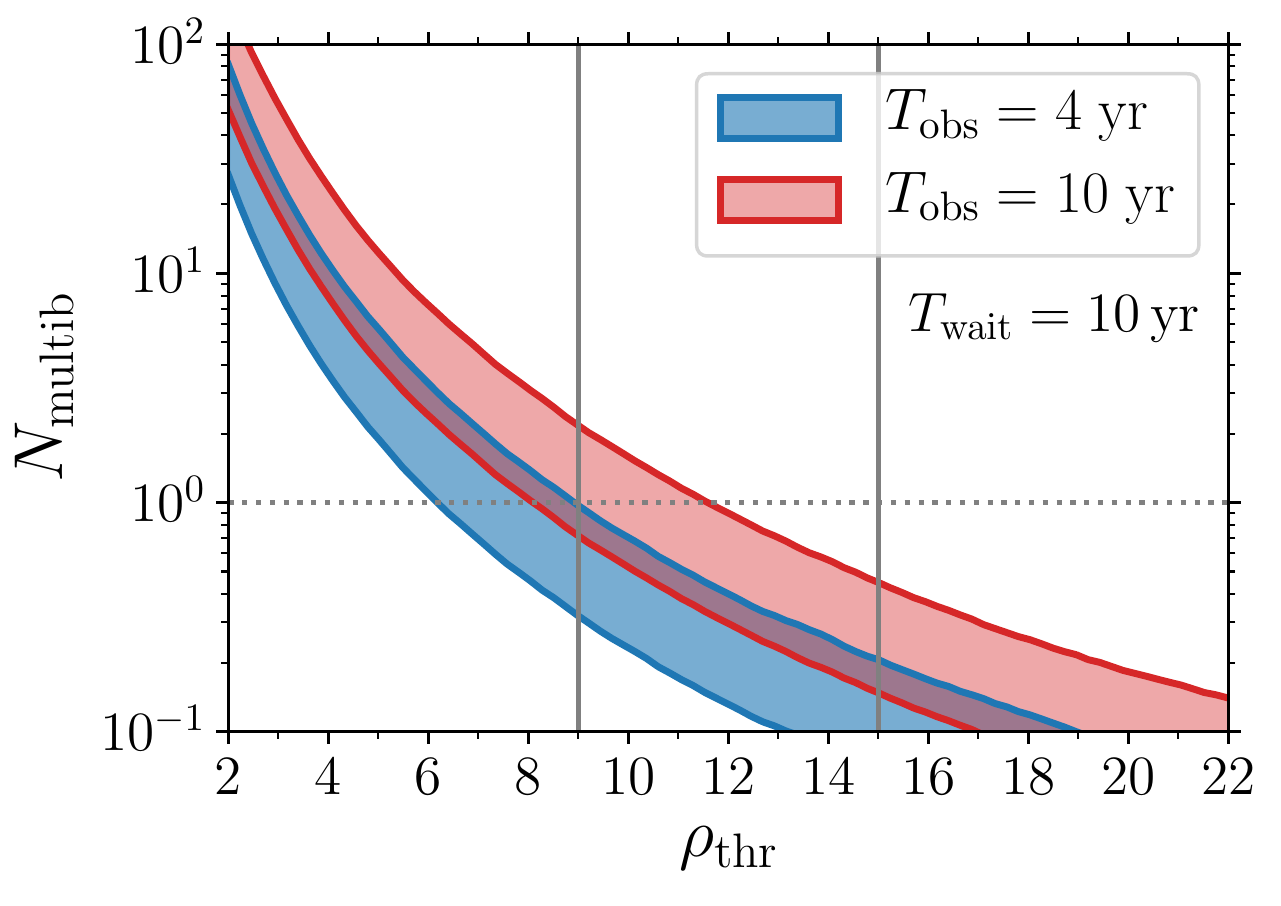}
        \caption{Number of stellar-mass BH binaries jointly detectable by the LISA space mission and ground-based interferometers as a function of the threshold SNR. Blue (red) curves assume a LISA mission duration $T_{\rm obs}=4\,$yr (10 yr). We only consider binaries merging within $T_{\rm wait}=10\,$yr. For each set of parameters, the shaded areas captures the current uncertainties in the local BH merger rate; in particular, we set $\mathcal{R}=97 \,{\rm Gpc}^{-3} {\rm yr}^{-1}$ ($32 \,{\rm Gpc}^{-3} {\rm yr}^{-1}$) for the upper (lower) line in each set. Vertical solid lines mark the SNR thresholds estimated in this letter for both forewarnings ($\rho_{\rm thr}\ssim 15$, c.f. Fig.~\ref{checkTwait}) and targeted archival searches ($\rho_{\rm thr}\ssim 9$). 
        }
        \label{impactSNRthr}
    \end{figure}

    Figure~\ref{impactSNRthr} shows the number of multiband detections merging within $T_{\rm wait}=10$~yr as a function of the SNR threshold. Multiband sources will be restricted to the the local Universe, where the probability distribution function of the SNRs assumes the universal form $p(\rho)\propto \rho^{-4}$ \citep{2011CQGra..28l5023S,2014arXiv1409.0522C}. The number of detections above threshold, therefore, scales as
    \begin{equation}
    N(\rho_{\rm thr})  \propto \int_{\rho > \rho_{\rm thr}} \frac{1}{\rho^4} \propto \frac{1}{\rho_{\rm thr}^3}\,.
    \label{Nvsrhothr}
    \end{equation}
    This severe scaling means that even a modest increase of the threshold SNR can push the number of sources below unity. Unfortunately, this turns out to be the case in most of these models. Using $\rho_{\rm thr}=15$, we predict LISA will not provide forewarnings to ground-based detectors for this population of stellar-mass BHs.
    
    As shown in Sec.~\ref{sec:Nbank}, archival searches require smaller template banks, lowering the SNR threshold to $\sim 9$. In this case, we find $0.5\lesssim N_{\rm multib}\lesssim 2$. Revisiting past LISA data, as first put forward by \cite{2018PhRvL.121y1102W}, might well be our only chance to observe stellar-mass BH binaries with LISA.

\begin{figure}
    \centering
        \includegraphics[width=0.45\textwidth]{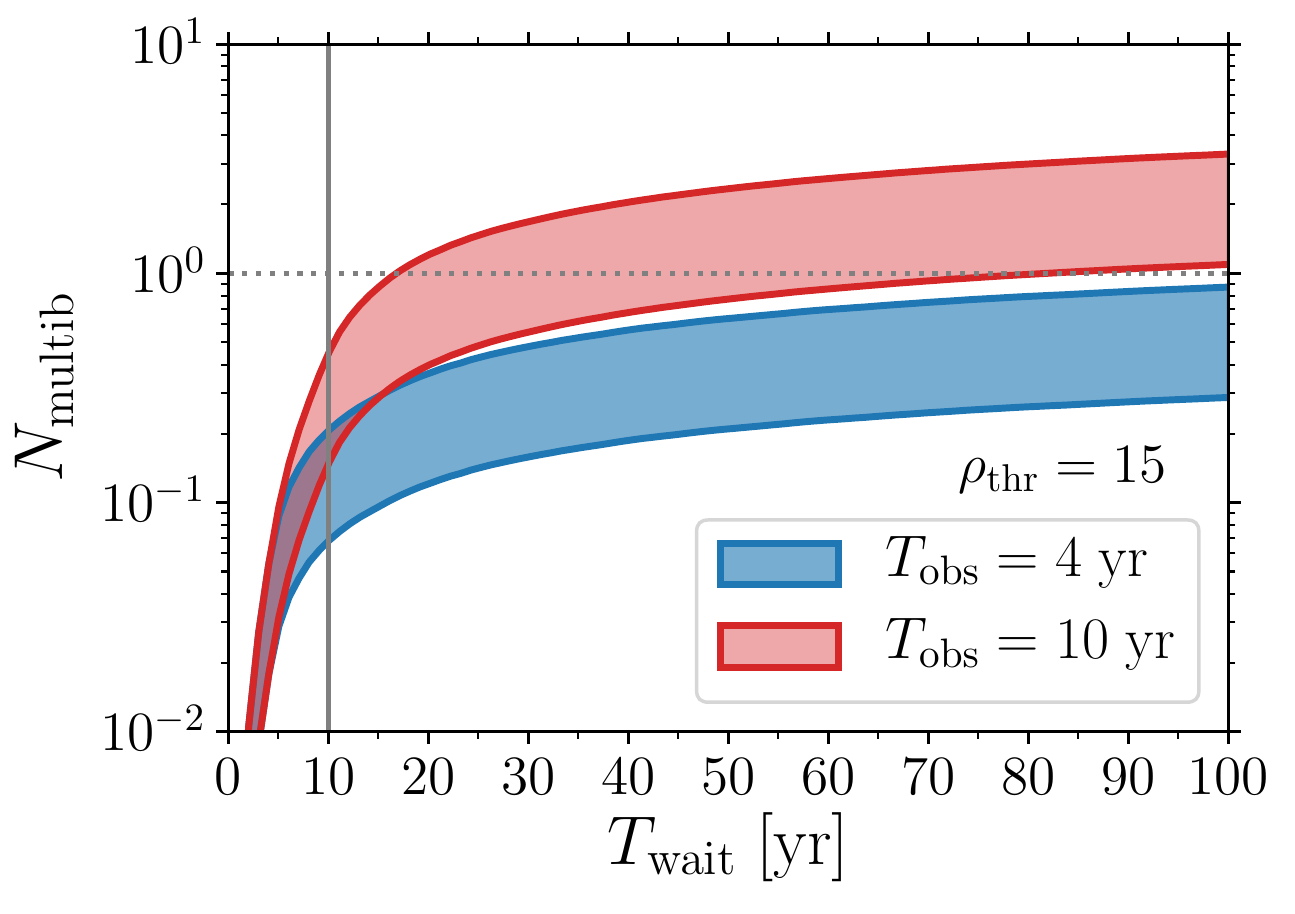}
        \caption{Number of stellar-mass BH binaries merging within $T_{\rm wait}$ observable by LISA with $\rho \geq 15$.  Blue (red) curves assume a LISA mission duration $T_{\rm obs}=4\,$yr (10 yr). For each set of parameters, the shaded areas captures the current uncertainties in the local BH merger rate; in particular, we set $\mathcal{R}=97 \,{\rm Gpc}^{-3} {\rm yr}^{-1}$ ($32 \,{\rm Gpc}^{-3} {\rm yr}^{-1}$) for the upper (lower) line in each set. The vertical line marks $T_{\rm wait}=10$yr, as used in Fig.~\ref{impactSNRthr}. 
        }
        \label{checkTwait}
    \end{figure}

    Some events from the population of binaries in the early inspiral might also be above threshold (Fig.~\ref{checkTwait}). If the mission is long enough, we find that a few sources with merger times $t_{\rm merger}\lesssim 100\,$yr will be observable by LISA. 
 For these slowly chirping signals, LISA will be able to provide forewarnings of a small number of events a very long time into the future.

    \section{Discussion and Conclusions}
    \label{sec:conclusions}
    
    In this letter we have considered the LISA detectability of BH binaries with component masses in the range $(5\!-\!50)\,M_\odot$.%
    We find that, due to the complexity of the signal space, an idealised template bank search has a threshold SNR of $\rho_{\rm thr}\approx14\!-\!15$. 
        This is significantly higher than previous assumed. Because the expected number of events scales as $\rho_{\rm thr}^{-3}$, our revised estimate implies that LISA might not provide forewarnings for any ground-based detectors within a 10~yr timescale.  
    From a data analysis perspective, stellar-mass BH binaries in LISA  are in some respects more similar to EMRIs than to LIGO/Virgo binary BHs. 
Our estimate applies to an optimal template-bank search; in practice, a sub-optimal approach may be required, further raising $\rho_{\rm thr}$. 
We stress that our calculation is only a preliminary estimate and will need to be corroborated with future injection campaigns.
    
    Because the expected numbers of events is so low, it is crucial to maximise our sensitivity to these events using all the tools at our disposal.
    Figure~\ref{checkTwait} shows the importance of  a long mission duration. With $T_{\rm}=4\,\mathrm{yrs}$, LISA might not be able to provide  forewarnings even 100~yrs into the future. Conversely, a 10 yr mission might deliver a few sources with $t_{\rm merger}\lesssim 100$ yr.
    These stellar-mass events exists at the high frequency end of the LISA sensitivity window. Therefore it is crucial to preserve the detector performance in this region. 
  It is possible the high-frequency noise level might turn out to be up to $\ssim 1.5$ times lower than assumed here \citep{2017arXiv170200786A}. 
Lowering the noise floor increases the SNR linearly, which in turn increases the expected number of events by a factor of up to $1.5^3\approx3.4$.

Detections from the ground can be used to dig deeper into archival LISA data. In this case, a smaller template bank will be sufficient, bringing the SNR threshold down to $\ssim 9$ and the expected number of detections up to a few.

In this paper we have considered BHs with merger times up to 100~yrs. A numerous population with larger merger times might also be present. The Milky Way alone could host millions of $\ssim 30 M_{\odot}$ BHs in wide orbits \citep{ 2018MNRAS.473.1186E,2018MNRAS.480.2704L}. 
    Because these systems are slowly chirping and closer to being monochromatic, the signal space is considerably simpler allowing for a lower threshold SNR.  We defer an analysis of this population to future work.
 
The rate estimates of Sec.~\ref{sec:implications} depends on the largest BH mass considered, here taken to be $50 M_\odot$. This value is motivated by current LIGO/Virgo observations as well as theoretical predictions of supernova instabilities~\citep{1967PhRvL..18..379B}. 
If, however, a population of BHs with masses $\ssim 100 M_{\odot}$ were to be present,
such systems would be prominent multiband sources. Their larger mass would imply both larger LISA SNRs and shorter merger times. Repeating the analysis of Sec.~\ref{sec:implications} with a cutoff of $100 M_{\odot}$ yields $0.5\lesssim N_{\rm multib} \lesssim 4$ for $\rho_{\rm thr}=15$ and $T_{\rm wait}=10$ yr.

Stellar-mass BHs binaries in the mHz regime are intrinsically quiet and their observation with LISA will be challenging. A combination of detector-sensitivity improvements, data-analysis advancements, and possibly a pinch of luck, might all turn out to be necessary.

        \vspace{-0.5cm}
    \section*{Acknowledgements}
    
    We thank A.\ Vecchio, A.\ J.\ K.\ Chua, P.\ McNamara, E.\ Berti, K.\ Wong and B.\ S.\ Sathyaprakash for discussions.
    Computational work was performed on the University of Birmingham's BlueBEAR cluster and at the Maryland Advanced Research Computing Center (MARCC).

        \newpage    

    \vspace{-0.5cm}
    \bibliographystyle{mnras_tex_edited.bst}
    \bibliography{refs} %

\begin{thebibliography}{}
\makeatletter
\relax
\def\mn@urlcharsother{\let\do\@makeother \do\$\do\&\do\#\do\^\do\_\do\%\do\~}
\def\mn@doi{\begingroup\mn@urlcharsother \@ifnextchar [ {\mn@doi@}
  {\mn@doi@[]}}
\def\mn@doi@[#1]#2{\def\@tempa{#1}\ifx\@tempa\@empty \href
  {http://dx.doi.org/#2} {doi:#2}\else \href {http://dx.doi.org/#2} {#1}\fi
  \endgroup}
\def\mn@eprint#1#2{\mn@eprint@#1:#2::\@nil}
\def\mn@eprint@arXiv#1{\href {http://arxiv.org/abs/#1} {{arXiv:#1}}}
\def\mn@eprint@dblp#1{\href {http://dblp.uni-trier.de/rec/bibtex/#1.xml}
  {dblp:#1}}
\def\mn@eprint@#1:#2:#3:#4\@nil{\def\@tempa {#1}\def\@tempb {#2}\def\@tempc
  {#3}\ifx \@tempc \@empty \let \@tempc \@tempb \let \@tempb \@tempa \fi \ifx
  \@tempb \@empty \def\@tempb {arXiv}\fi \@ifundefined
  {mn@eprint@\@tempb}{\@tempb:\@tempc}{\expandafter \expandafter \csname
  mn@eprint@\@tempb\endcsname \expandafter{\@tempc}}}

\bibitem[\protect\citeauthoryear{{Abbott} et~al.,}{{Abbott}
  et~al.}{2016a}]{2016PhRvX...6d1015A}
{Abbott} B.~P.,  et~al., 2016a, \mn@doi [\prx] {10.1103/PhysRevX.6.041015},
  \href {https://ui.adsabs.harvard.edu/abs/2016PhRvX...6d1015A} {6, 041015}
  (\mn@eprint {arXiv} {1606.04856})

\bibitem[\protect\citeauthoryear{{Abbott} et~al.,}{{Abbott}
  et~al.}{2016b}]{2016PhRvL.116x1102A}
{Abbott} B.~P.,  et~al., 2016b, \mn@doi [\prl]
  {10.1103/PhysRevLett.116.241102}, \href
  {https://ui.adsabs.harvard.edu/abs/2016PhRvL.116x1102A} {116, 241102}
  (\mn@eprint {arXiv} {1602.03840})

\bibitem[\protect\citeauthoryear{{Abbott} et~al.}{{Abbott}
  et~al.}{2016c}]{2016ApJ...818L..22A}
{Abbott} B.~P.,  et~al., 2016c, \mn@doi [\apjl] {10.3847/2041-8205/818/2/L22},
  \href {https://ui.adsabs.harvard.edu/abs/2016ApJ...818L..22A} {818, L22}
  (\mn@eprint {arXiv} {1602.03846})

\bibitem[\protect\citeauthoryear{{Abbott} et~al.}{{Abbott}
  et~al.}{2016d}]{2016ApJ...833L...1A}
{Abbott} B.~P.,  et~al., 2016d, \mn@doi [\apjl] {10.3847/2041-8205/833/1/L1},
  \href {https://ui.adsabs.harvard.edu/abs/2016ApJ...833L...1A} {833, L1}
  (\mn@eprint {arXiv} {1602.03842})

\bibitem[\protect\citeauthoryear{{Abbott} et~al.}{{Abbott}
  et~al.}{2018a}]{2018arXiv181112907T}
{Abbott} B.~P.,  et~al., 2018a, \href
  {https://ui.adsabs.harvard.edu/abs/2018arXiv181112907T} {} (\mn@eprint
  {arXiv} {1811.12907})

\bibitem[\protect\citeauthoryear{{Abbott} et~al.}{{Abbott}
  et~al.}{2018b}]{2018LRR....21....3A}
{Abbott} B.~P.,  et~al., 2018b, \mn@doi [\lrr] {10.1007/s41114-018-0012-9},
  \href {https://ui.adsabs.harvard.edu/abs/2018LRR....21....3A} {21, 3}
  (\mn@eprint {arXiv} {1304.0670})

\bibitem[\protect\citeauthoryear{{Ade} et~al.}{{Ade}
  et~al.}{2016}]{2016A&A...594A..13P}
{Ade} P.~A.~R.,  et~al., 2016, \mn@doi [\aap] {10.1051/0004-6361/201525830},
  \href {https://ui.adsabs.harvard.edu/abs/2016A%26A...594A..13P} {594, A13}
  (\mn@eprint {arXiv} {1502.01589})

\bibitem[\protect\citeauthoryear{{Amaro-Seoane} et~al.,}{{Amaro-Seoane}
  et~al.}{2017}]{2017arXiv170200786A}
{Amaro-Seoane} P.,  et~al., 2017, \href
  {https://ui.adsabs.harvard.edu/abs/2017arXiv170200786A} {} (\mn@eprint
  {arXiv} {1702.00786})

\bibitem[\protect\citeauthoryear{{Barausse}, {Yunes}  \&
  {Chamberlain}}{{Barausse} et~al.}{2016}]{2016PhRvL.116x1104B}
{Barausse} E.,  {Yunes} N.,   {Chamberlain} K.,  2016, \mn@doi [\prl]
  {10.1103/PhysRevLett.116.241104}, \href
  {https://ui.adsabs.harvard.edu/abs/2016PhRvL.116x1104B} {116, 241104}
  (\mn@eprint {arXiv} {1603.04075})

\bibitem[\protect\citeauthoryear{{Barkat}, {Rakavy}  \& {Sack}}{{Barkat}
  et~al.}{1967}]{1967PhRvL..18..379B}
{Barkat} Z.,  {Rakavy} G.,   {Sack} N.,  1967, \mn@doi [\prl]
  {10.1103/PhysRevLett.18.379}, \href
  {https://ui.adsabs.harvard.edu/abs/1967PhRvL..18..379B} {18, 379}

\bibitem[\protect\citeauthoryear{{Belczynski}, {Bulik}, {Fryer}, {Ruiter},
  {Valsecchi}, {Vink}  \& {Hurley}}{{Belczynski}
  et~al.}{2010a}]{2010ApJ...714.1217B}
{Belczynski} K.,  {Bulik} T.,  {Fryer} C.~L.,  {Ruiter} A.,  {Valsecchi} F.,
  {Vink} J.~S.,   {Hurley} J.~R.,  2010a, \mn@doi [\apj]
  {10.1088/0004-637X/714/2/1217}, \href
  {https://ui.adsabs.harvard.edu/abs/2010ApJ...714.1217B} {714, 1217}
  (\mn@eprint {arXiv} {0904.2784})

\bibitem[\protect\citeauthoryear{{Belczynski}, {Benacquista}  \&
  {Bulik}}{{Belczynski} et~al.}{2010b}]{2010ApJ...725..816B}
{Belczynski} K.,  {Benacquista} M.,   {Bulik} T.,  2010b, \mn@doi [\apj]
  {10.1088/0004-637X/725/1/816}, \href
  {https://ui.adsabs.harvard.edu/abs/2010ApJ...725..816B} {725, 816}
  (\mn@eprint {arXiv} {0811.1602})

\bibitem[\protect\citeauthoryear{{Brady}, {Creighton}, {Cutler}  \&
  {Schutz}}{{Brady} et~al.}{1998}]{1998PhRvD..57.2101B}
{Brady} P.~R.,  {Creighton} T.,  {Cutler} C.,   {Schutz} B.~F.,  1998, \mn@doi
  [\prd] {10.1103/PhysRevD.57.2101}, \href
  {https://ui.adsabs.harvard.edu/abs/1998PhRvD..57.2101B} {57, 2101}
  (\mn@eprint {} {gr-qc/9702050})

\bibitem[\protect\citeauthoryear{{Breivik}, {Rodriguez}, {Larson}, {Kalogera}
  \& {Rasio}}{{Breivik} et~al.}{2016}]{2016ApJ...830L..18B}
{Breivik} K.,  {Rodriguez} C.~L.,  {Larson} S.~L.,  {Kalogera} V.,   {Rasio}
  F.~A.,  2016, \mn@doi [\apjl] {10.3847/2041-8205/830/1/L18}, \href
  {https://ui.adsabs.harvard.edu/abs/2016ApJ...830L..18B} {830, L18}
  (\mn@eprint {arXiv} {1606.09558})

\bibitem[\protect\citeauthoryear{{Buonanno}, {Chen}  \&
  {Vallisneri}}{{Buonanno} et~al.}{2003}]{2003PhRvD..67b4016B}
{Buonanno} A.,  {Chen} Y.,   {Vallisneri} M.,  2003, \mn@doi [\prd]
  {10.1103/PhysRevD.67.024016}, \href
  {https://ui.adsabs.harvard.edu/abs/2003PhRvD..67b4016B} {67, 024016}
  (\mn@eprint {} {gr-qc/0205122})

\bibitem[\protect\citeauthoryear{{Cannon}, {Chapman}, {Hanna}, {Keppel},
  {Searle}  \& {Weinstein}}{{Cannon} et~al.}{2010}]{2010PhRvD..82d4025C}
{Cannon} K.,  {Chapman} A.,  {Hanna} C.,  {Keppel} D.,  {Searle} A.~C.,
  {Weinstein} A.~J.,  2010, \mn@doi [\prd] {10.1103/PhysRevD.82.044025}, \href
  {https://ui.adsabs.harvard.edu/abs/2010PhRvD..82d4025C} {82, 044025}
  (\mn@eprint {arXiv} {1005.0012})

\bibitem[\protect\citeauthoryear{{Chamberlain} \& {Yunes}}{{Chamberlain} \&
  {Yunes}}{2017}]{2017PhRvD..96h4039C}
{Chamberlain} K.,  {Yunes} N.,  2017, \mn@doi [\prd]
  {10.1103/PhysRevD.96.084039}, \href
  {https://ui.adsabs.harvard.edu/abs/2017PhRvD..96h4039C} {96, 084039}
  (\mn@eprint {arXiv} {1704.08268})

\bibitem[\protect\citeauthoryear{{Chen} \& {Holz}}{{Chen} \&
  {Holz}}{2014}]{2014arXiv1409.0522C}
{Chen} H.-Y.,  {Holz} D.~E.,  2014, \href
  {https://ui.adsabs.harvard.edu/abs/2014arXiv1409.0522C} {} (\mn@eprint
  {arXiv} {1409.0522})

\bibitem[\protect\citeauthoryear{{Chua}, {Moore}  \& {Gair}}{{Chua}
  et~al.}{2017}]{2017PhRvD..96d4005C}
{Chua} A.~J.~K.,  {Moore} C.~J.,   {Gair} J.~R.,  2017, \mn@doi [\prd]
  {10.1103/PhysRevD.96.044005}, \href
  {https://ui.adsabs.harvard.edu/abs/2017PhRvD..96d4005C} {96, 044005}
  (\mn@eprint {arXiv} {1705.04259})

\bibitem[\protect\citeauthoryear{{Cornish} \& {Porter}}{{Cornish} \&
  {Porter}}{2005}]{2005CQGra..22S.927C}
{Cornish} N.~J.,  {Porter} E.~K.,  2005, \mn@doi [\cqg]
  {10.1088/0264-9381/22/18/S06}, \href
  {https://ui.adsabs.harvard.edu/abs/2005CQGra..22S.927C} {22, S927}
  (\mn@eprint {} {gr-qc/0504012})

\bibitem[\protect\citeauthoryear{{Cutler}}{{Cutler}}{1998}]{1998PhRvD..57.7089C}
{Cutler} C.,  1998, \mn@doi [\prd] {10.1103/PhysRevD.57.7089}, \href
  {https://ui.adsabs.harvard.edu/abs/1998PhRvD..57.7089C} {57, 7089}
  (\mn@eprint {} {gr-qc/9703068})

\bibitem[\protect\citeauthoryear{{Cutler} et~al.,}{{Cutler}
  et~al.}{2019}]{2019BAAS...51c.109C}
{Cutler} C.,  et~al., 2019, Bulletin of the American Astronomical Society,
  \href {https://ui.adsabs.harvard.edu/abs/2019BAAS...51c.109C} {51, 109}
  (\mn@eprint {arXiv} {1903.04069})

\bibitem[\protect\citeauthoryear{{D'Orazio} \& {Samsing}}{{D'Orazio} \&
  {Samsing}}{2018}]{2018MNRAS.481.4775D}
{D'Orazio} D.~J.,  {Samsing} J.,  2018, \mn@doi [\mnras]
  {10.1093/mnras/sty2568}, \href
  {https://ui.adsabs.harvard.edu/abs/2018MNRAS.481.4775D} {481, 4775}
  (\mn@eprint {arXiv} {1805.06194})

\bibitem[\protect\citeauthoryear{{Dal Canton} \& {Harry}}{{Dal Canton} \&
  {Harry}}{2017}]{2017arXiv170501845D}
{Dal Canton} T.,  {Harry} I.~W.,  2017, \href
  {https://ui.adsabs.harvard.edu/abs/2017arXiv170501845D} {} (\mn@eprint
  {arXiv} {1705.01845})

\bibitem[\protect\citeauthoryear{{Del Pozzo}, {Sesana}  \& {Klein}}{{Del Pozzo}
  et~al.}{2018}]{2018MNRAS.475.3485D}
{Del Pozzo} W.,  {Sesana} A.,   {Klein} A.,  2018, \mn@doi [\mnras]
  {10.1093/mnras/sty057}, \href
  {https://ui.adsabs.harvard.edu/abs/2018MNRAS.475.3485D} {475, 3485}
  (\mn@eprint {arXiv} {1703.01300})

\bibitem[\protect\citeauthoryear{{Dhurandhar} \& {Sathyaprakash}}{{Dhurandhar}
  \& {Sathyaprakash}}{1994}]{1994PhRvD..49.1707D}
{Dhurandhar} S.~V.,  {Sathyaprakash} B.~S.,  1994, \mn@doi [\prd]
  {10.1103/PhysRevD.49.1707}, \href
  {https://ui.adsabs.harvard.edu/abs/1994PhRvD..49.1707D} {49, 1707}

\bibitem[\protect\citeauthoryear{{Elbert}, {Bullock}  \& {Kaplinghat}}{{Elbert}
  et~al.}{2018}]{2018MNRAS.473.1186E}
{Elbert} O.~D.,  {Bullock} J.~S.,   {Kaplinghat} M.,  2018, \mn@doi [\mnras]
  {10.1093/mnras/stx1959}, \href
  {https://ui.adsabs.harvard.edu/abs/2018MNRAS.473.1186E} {473, 1186}
  (\mn@eprint {arXiv} {1703.02551})

\bibitem[\protect\citeauthoryear{{Fang}, {Thompson}  \& {Hirata}}{{Fang}
  et~al.}{2019}]{2019ApJ...875...75F}
{Fang} X.,  {Thompson} T.~A.,   {Hirata} C.~M.,  2019, \mn@doi [\apj]
  {10.3847/1538-4357/ab0e6a}, \href
  {https://ui.adsabs.harvard.edu/abs/2019ApJ...875...75F} {875, 75} (\mn@eprint
  {arXiv} {1901.05092})

\bibitem[\protect\citeauthoryear{{Field}, {Galley}, {Herrmann}, {Hesthaven},
  {Ochsner}  \& {Tiglio}}{{Field} et~al.}{2011}]{2011PhRvL.106v1102F}
{Field} S.~E.,  {Galley} C.~R.,  {Herrmann} F.,  {Hesthaven} J.~S.,  {Ochsner}
  E.,   {Tiglio} M.,  2011, \mn@doi [\prl] {10.1103/PhysRevLett.106.221102},
  \href {https://ui.adsabs.harvard.edu/abs/2011PhRvL.106v1102F} {106, 221102}
  (\mn@eprint {arXiv} {1101.3765})

\bibitem[\protect\citeauthoryear{{Finn} \& {Chernoff}}{{Finn} \&
  {Chernoff}}{1993}]{1993PhRvD..47.2198F}
{Finn} L.~S.,  {Chernoff} D.~F.,  1993, \mn@doi [\prd]
  {10.1103/PhysRevD.47.2198}, \href
  {https://ui.adsabs.harvard.edu/abs/1993PhRvD..47.2198F} {47, 2198}
  (\mn@eprint {} {gr-qc/9301003})

\bibitem[\protect\citeauthoryear{{Gair}, {Barack}, {Creighton}, {Cutler},
  {Larson}, {Phinney}  \& {Vallisneri}}{{Gair}
  et~al.}{2004}]{2004CQGra..21S1595G}
{Gair} J.~R.,  {Barack} L.,  {Creighton} T.,  {Cutler} C.,  {Larson} S.~L.,
  {Phinney} E.~S.,   {Vallisneri} M.,  2004, \mn@doi [\cqg]
  {10.1088/0264-9381/21/20/003}, \href
  {https://ui.adsabs.harvard.edu/abs/2004CQGra..21S1595G} {21, S1595}
  (\mn@eprint {} {gr-qc/0405137})

\bibitem[\protect\citeauthoryear{{Gerosa}, {Ma}, {Wong}, {Berti},
  {O'Shaughnessy}, {Chen}  \& {Belczynski}}{{Gerosa}
  et~al.}{2019}]{2019PhRvD..99j3004G}
{Gerosa} D.,  {Ma} S.,  {Wong} K.~W.~K.,  {Berti} E.,  {O'Shaughnessy} R.,
  {Chen} Y.,   {Belczynski} K.,  2019, \mn@doi [\prd]
  {10.1103/PhysRevD.99.103004}, \href
  {https://ui.adsabs.harvard.edu/abs/2019PhRvD..99j3004G} {99, 103004}
  (\mn@eprint {arXiv} {1902.00021})

\bibitem[\protect\citeauthoryear{{Klein}, {Boetzel}, {Gopakumar}, {Jetzer}  \&
  {de Vittori}}{{Klein} et~al.}{2018}]{2018PhRvD..98j4043K}
{Klein} A.,  {Boetzel} Y.,  {Gopakumar} A.,  {Jetzer} P.,   {de Vittori} L.,
  2018, \mn@doi [\prd] {10.1103/PhysRevD.98.104043}, \href
  {https://ui.adsabs.harvard.edu/abs/2018PhRvD..98j4043K} {98, 104043}
  (\mn@eprint {arXiv} {1801.08542})

\bibitem[\protect\citeauthoryear{{Kremer}, {Chatterjee}, {Breivik},
  {Rodriguez}, {Larson}  \& {Rasio}}{{Kremer}
  et~al.}{2018}]{2018PhRvL.120s1103K}
{Kremer} K.,  {Chatterjee} S.,  {Breivik} K.,  {Rodriguez} C.~L.,  {Larson}
  S.~L.,   {Rasio} F.~A.,  2018, \mn@doi [\prl]
  {10.1103/PhysRevLett.120.191103}, \href
  {https://ui.adsabs.harvard.edu/abs/2018PhRvL.120s1103K} {120, 191103}
  (\mn@eprint {arXiv} {1802.05661})

\bibitem[\protect\citeauthoryear{{Kremer} et~al.,}{{Kremer}
  et~al.}{2019}]{2019PhRvD..99f3003K}
{Kremer} K.,  et~al., 2019, \mn@doi [\prd] {10.1103/PhysRevD.99.063003}, \href
  {https://ui.adsabs.harvard.edu/abs/2019PhRvD..99f3003K} {99, 063003}
  (\mn@eprint {arXiv} {1811.11812})

\bibitem[\protect\citeauthoryear{{Kyutoku} \& {Seto}}{{Kyutoku} \&
  {Seto}}{2016}]{2016MNRAS.462.2177K}
{Kyutoku} K.,  {Seto} N.,  2016, \mn@doi [\mnras] {10.1093/mnras/stw1767},
  \href {https://ui.adsabs.harvard.edu/abs/2016MNRAS.462.2177K} {462, 2177}
  (\mn@eprint {arXiv} {1606.02298})

\bibitem[\protect\citeauthoryear{{Kyutoku} \& {Seto}}{{Kyutoku} \&
  {Seto}}{2017}]{2017PhRvD..95h3525K}
{Kyutoku} K.,  {Seto} N.,  2017, \mn@doi [\prd] {10.1103/PhysRevD.95.083525},
  \href {https://ui.adsabs.harvard.edu/abs/2017PhRvD..95h3525K} {95, 083525}
  (\mn@eprint {arXiv} {1609.07142})

\bibitem[\protect\citeauthoryear{{Lamberts} et~al.,}{{Lamberts}
  et~al.}{2018}]{2018MNRAS.480.2704L}
{Lamberts} A.,  et~al., 2018, \mn@doi [\mnras] {10.1093/mnras/sty2035}, \href
  {https://ui.adsabs.harvard.edu/abs/2018MNRAS.480.2704L} {480, 2704}
  (\mn@eprint {arXiv} {1801.03099})

\bibitem[\protect\citeauthoryear{{Mangiagli}, {Klein}, {Sesana}, {Barausse}  \&
  {Colpi}}{{Mangiagli} et~al.}{2019}]{2019PhRvD..99f4056M}
{Mangiagli} A.,  {Klein} A.,  {Sesana} A.,  {Barausse} E.,   {Colpi} M.,  2019,
  \mn@doi [\prd] {10.1103/PhysRevD.99.064056}, \href
  {https://ui.adsabs.harvard.edu/abs/2019PhRvD..99f4056M} {99, 064056}
  (\mn@eprint {arXiv} {1811.01805})

\bibitem[\protect\citeauthoryear{{Messick} et~al.,}{{Messick}
  et~al.}{2017}]{2017PhRvD..95d2001M}
{Messick} C.,  et~al., 2017, \mn@doi [\prd] {10.1103/PhysRevD.95.042001}, \href
  {https://ui.adsabs.harvard.edu/abs/2017PhRvD..95d2001M} {95, 042001}
  (\mn@eprint {arXiv} {1604.04324})

\bibitem[\protect\citeauthoryear{{Nishizawa}, {Berti}, {Klein}  \&
  {Sesana}}{{Nishizawa} et~al.}{2016}]{2016PhRvD..94f4020N}
{Nishizawa} A.,  {Berti} E.,  {Klein} A.,   {Sesana} A.,  2016, \mn@doi [\prd]
  {10.1103/PhysRevD.94.064020}, \href
  {https://ui.adsabs.harvard.edu/abs/2016PhRvD..94f4020N} {94, 064020}
  (\mn@eprint {arXiv} {1605.01341})

\bibitem[\protect\citeauthoryear{{Nishizawa}, {Sesana}, {Berti}  \&
  {Klein}}{{Nishizawa} et~al.}{2017}]{2017MNRAS.465.4375N}
{Nishizawa} A.,  {Sesana} A.,  {Berti} E.,   {Klein} A.,  2017, \mn@doi
  [\mnras] {10.1093/mnras/stw2993}, \href
  {https://ui.adsabs.harvard.edu/abs/2017MNRAS.465.4375N} {465, 4375}
  (\mn@eprint {arXiv} {1606.09295})

\bibitem[\protect\citeauthoryear{{Owen}}{{Owen}}{1996}]{1996PhRvD..53.6749O}
{Owen} B.~J.,  1996, \mn@doi [\prd] {10.1103/PhysRevD.53.6749}, \href
  {https://ui.adsabs.harvard.edu/abs/1996PhRvD..53.6749O} {53, 6749}
  (\mn@eprint {} {gr-qc/9511032})

\bibitem[\protect\citeauthoryear{{Owen} \& {Sathyaprakash}}{{Owen} \&
  {Sathyaprakash}}{1999}]{1999PhRvD..60b2002O}
{Owen} B.~J.,  {Sathyaprakash} B.~S.,  1999, \mn@doi [\prd]
  {10.1103/PhysRevD.60.022002}, \href
  {https://ui.adsabs.harvard.edu/abs/1999PhRvD..60b2002O} {60, 022002}
  (\mn@eprint {} {gr-qc/9808076})

\bibitem[\protect\citeauthoryear{{Randall} \& {Xianyu}}{{Randall} \&
  {Xianyu}}{2019}]{2019ApJ...878...75R}
{Randall} L.,  {Xianyu} Z.-Z.,  2019, \mn@doi [\apj]
  {10.3847/1538-4357/ab20c6}, \href
  {https://ui.adsabs.harvard.edu/abs/2019ApJ...878...75R} {878, 75} (\mn@eprint
  {arXiv} {1805.05335})

\bibitem[\protect\citeauthoryear{{Robson}, {Cornish}  \& {Liug}}{{Robson}
  et~al.}{2019}]{2019CQGra..36j5011R}
{Robson} T.,  {Cornish} N.~J.,   {Liug} C.,  2019, \mn@doi [\cqg]
  {10.1088/1361-6382/ab1101}, \href
  {https://ui.adsabs.harvard.edu/abs/2019CQGra..36j5011R} {36, 105011}
  (\mn@eprint {arXiv} {1803.01944})

\bibitem[\protect\citeauthoryear{{Samsing} \& {D'Orazio}}{{Samsing} \&
  {D'Orazio}}{2018}]{2018MNRAS.481.5445S}
{Samsing} J.,  {D'Orazio} D.~J.,  2018, \mn@doi [\mnras]
  {10.1093/mnras/sty2334}, \href
  {https://ui.adsabs.harvard.edu/abs/2018MNRAS.481.5445S} {481, 5445}
  (\mn@eprint {arXiv} {1804.06519})

\bibitem[\protect\citeauthoryear{{Samsing} \& {D'Orazio}}{{Samsing} \&
  {D'Orazio}}{2019}]{2019PhRvD..99f3006S}
{Samsing} J.,  {D'Orazio} D.~J.,  2019, \mn@doi [\prd]
  {10.1103/PhysRevD.99.063006}, \href
  {https://ui.adsabs.harvard.edu/abs/2019PhRvD..99f3006S} {99, 063006}
  (\mn@eprint {arXiv} {1807.08864})

\bibitem[\protect\citeauthoryear{{Samsing}, {D'Orazio}, {Askar}  \&
  {Giersz}}{{Samsing} et~al.}{2018}]{2018arXiv180208654S}
{Samsing} J.,  {D'Orazio} D.~J.,  {Askar} A.,   {Giersz} M.,  2018, \href
  {https://ui.adsabs.harvard.edu/abs/2018arXiv180208654S} {} (\mn@eprint
  {arXiv} {1802.08654})

\bibitem[\protect\citeauthoryear{{Santamar{\'{\i}}a}
  et~al.,}{{Santamar{\'{\i}}a} et~al.}{2010}]{2010PhRvD..82f4016S}
{Santamar{\'{\i}}a} L.,  et~al., 2010, \mn@doi [\prd]
  {10.1103/PhysRevD.82.064016}, \href
  {https://ui.adsabs.harvard.edu/abs/2010PhRvD..82f4016S} {82, 064016}
  (\mn@eprint {arXiv} {1005.3306})

\bibitem[\protect\citeauthoryear{{Sathyaprakash} \&
  {Dhurandhar}}{{Sathyaprakash} \& {Dhurandhar}}{1991}]{1991PhRvD..44.3819S}
{Sathyaprakash} B.~S.,  {Dhurandhar} S.~V.,  1991, \mn@doi [\prd]
  {10.1103/PhysRevD.44.3819}, \href
  {https://ui.adsabs.harvard.edu/abs/1991PhRvD..44.3819S} {44, 3819}

\bibitem[\protect\citeauthoryear{{Schutz}}{{Schutz}}{2011}]{2011CQGra..28l5023S}
{Schutz} B.~F.,  2011, \mn@doi [\cqg] {10.1088/0264-9381/28/12/125023}, \href
  {https://ui.adsabs.harvard.edu/abs/2011CQGra..28l5023S} {28, 125023}
  (\mn@eprint {arXiv} {1102.5421})

\bibitem[\protect\citeauthoryear{{Sesana}}{{Sesana}}{2016}]{2016PhRvL.116w1102S}
{Sesana} A.,  2016, \mn@doi [\prl] {10.1103/PhysRevLett.116.231102}, \href
  {https://ui.adsabs.harvard.edu/abs/2016PhRvL.116w1102S} {116, 231102}
  (\mn@eprint {arXiv} {1602.06951})

\bibitem[\protect\citeauthoryear{{Sesana}}{{Sesana}}{2017}]{2017JPhCS.840a2018S}
{Sesana} A.,  2017, \mn@doi [JPCS] {10.1088/1742-6596/840/1/012018}, \href
  {https://ui.adsabs.harvard.edu/abs/2017JPhCS.840a2018S} {840, 012018}
  (\mn@eprint {arXiv} {1702.04356})

\bibitem[\protect\citeauthoryear{{Tanay}, {Klein}, {Berti}  \&
  {Nishizawa}}{{Tanay} et~al.}{2019}]{2019arXiv190508811T}
{Tanay} S.,  {Klein} A.,  {Berti} E.,   {Nishizawa} A.,  2019, \href
  {https://ui.adsabs.harvard.edu/abs/2019arXiv190508811T} {} (\mn@eprint
  {arXiv} {1905.08811})

\bibitem[\protect\citeauthoryear{{Tso}, {Gerosa}  \& {Chen}}{{Tso}
  et~al.}{2018}]{2018arXiv180700075T}
{Tso} R.,  {Gerosa} D.,   {Chen} Y.,  2018, \mn@doi [\prd]
  {10.1103/PhysRevD.99.124043}, \href
  {https://ui.adsabs.harvard.edu/abs/2018arXiv180700075T} {99, 124043}
  (\mn@eprint {arXiv} {1807.00075})

\bibitem[\protect\citeauthoryear{{Usman} et~al.,}{{Usman}
  et~al.}{2016}]{2016CQGra..33u5004U}
{Usman} S.~A.,  et~al., 2016, \mn@doi [\cqg] {10.1088/0264-9381/33/21/215004},
  \href {https://ui.adsabs.harvard.edu/abs/2016CQGra..33u5004U} {33, 215004}
  (\mn@eprint {arXiv} {1508.02357})

\bibitem[\protect\citeauthoryear{{Vitale}}{{Vitale}}{2016}]{2016PhRvL.117e1102V}
{Vitale} S.,  2016, \mn@doi [\prl] {10.1103/PhysRevLett.117.051102}, \href
  {https://ui.adsabs.harvard.edu/abs/2016PhRvL.117e1102V} {117, 051102}
  (\mn@eprint {arXiv} {1605.01037})

\bibitem[\protect\citeauthoryear{{Wiktorowicz}, {Belczynski}  \&
  {Maccarone}}{{Wiktorowicz} et~al.}{2014}]{2014bsee.confE..37W}
{Wiktorowicz} G.,  {Belczynski} K.,   {Maccarone} T.,  2014, \href
  {https://ui.adsabs.harvard.edu/abs/2014bsee.confE..37W} {} (\mn@eprint
  {arXiv} {1312.5924})

\bibitem[\protect\citeauthoryear{{Wong}, {Kovetz}, {Cutler}  \& {Berti}}{{Wong}
  et~al.}{2018}]{2018PhRvL.121y1102W}
{Wong} K.~W.~K.,  {Kovetz} E.~D.,  {Cutler} C.,   {Berti} E.,  2018, \mn@doi
  [\prl] {10.1103/PhysRevLett.121.251102}, \href
  {https://ui.adsabs.harvard.edu/abs/2018PhRvL.121y1102W} {121, 251102}
  (\mn@eprint {arXiv} {1808.08247})

\makeatother
\end{thebibliography}

    \label{lastpage}
\end{document}